\documentclass[preprint,authoryear,12pt]{elsarticle}

\usepackage{amssymb}

\usepackage{pslatex}
\usepackage{graphicx}
\usepackage{lineno}
\newcommand{\apj}{Astrophys. J.}
\newcommand{\aap}{Astron. Astrophys.}
\newcommand{\jgr}{J. Geophys. Res.}
\newcommand{\grl}{Geophys. Res. Lett.}
\newcommand{\planss}{Planet. Space Sci.}
\newcommand{\ssr}{Space Sci. Rev.}
\newcommand{\icarus}{Icarus}

\newcommand{\lya}{{Lyman-$\alpha$}}

\newcommand{\arcdeg}{\mbox{$^\circ$}}%

\newcommand{\lro}{{\it LRO}}

\def\Heone{He\,{\sc i}}
\def\lam{$\lambda$}

\journal{Icarus}

\begin{document}

\begin{frontmatter}



\title{Temporal Variability of Lunar Exospheric Helium During January 2012 from LRO/LAMP}


\author[jhu]{Paul D. Feldman\corref{cor1}}
\cortext[cor1]{Corresponding author}
\ead{pdf@pha.jhu.edu}

\author[apl]{Dana M. Hurley}

\author[swsa]{Kurt D. Retherford}

\author[swsa]{G. Randall Gladstone}

\author[swb]{S. Alan Stern}

\author[caz]{Wayne Pryor}

\author[swb]{Joel Wm. Parker}

\author[swb]{David E. Kaufmann}

\author[swsa]{Michael W. Davis}

\author[swsa]{Maarten H. Versteeg}

\author[]{LAMP team}

\address[jhu]{Johns Hopkins University, Department of Physics and
Astronomy, 3100 N. Charles Street, Baltimore, MD 21218 USA}

\address[apl]{Johns Hopkins University Applied Physics Laboratory,
Space Department,
11100 Johns Hopkins Road,
Laurel, MD 20723-6099 USA}

\address[swsa]{Southwest Research Institute, P. O. Drawer 28510, 
San Antonio, TX 78228-0510 USA}

\address[swb]{Southwest Research Institute, 
Department of Space Studies, Suite 300, 
1050 Walnut Street, 
Boulder, CO 80302-5150 USA}

\address[caz]{Central Arizona College, Coolidge, AZ 85128}

\begin{abstract}

We report observations of the lunar helium exosphere made between December 29, 2011, and January 26, 2012, with the Lyman Alpha Mapping Project (LAMP) ultraviolet spectrograph on NASA's {\it Lunar Reconnaissance Orbiter Mission} (\lro).  The observations were made of resonantly scattered \Heone\ \lam584 from illuminated atmosphere against the dark lunar surface on the dawn side of the terminator.  We find no or little variation of the derived surface He density with latitude but day-to-day variations that likely reflect variations in the solar wind alpha flux.  The 5-day passage of the Moon through the Earth's magnetotail results in a factor of two decrease in surface density, which is well explained by model simulations.

\end{abstract}


\end{frontmatter}


\section{Introduction}
\label{intro}

We have recently reported on the detection of helium in the lunar
exosphere using the Lyman Alpha Mapping Project (LAMP) ultraviolet
spectrograph on NASA's {\it Lunar Reconnaissance Orbiter Mission} (\lro)
\citep{Stern:2012}. These were the first measurements of He since the {\it in situ} Lunar Atmosphere Composition Experiment (LACE) measurements made following the {\it Apollo 17} mission in 1972 \citep{Hodges:1973,Hodges:1974} and the first time that He has been seen by remote sensing instrumentation.  LAMP measured resonantly scattered solar \Heone\ \lam584\ from above the sunlit lunar limb.  To correct for the interplanetary \Heone\ \lam584, the spacecraft
executed a slew to point LAMP at fixed celestial coordinates keeping the sky background constant during the maneuver.  These observations provided a
snapshot of the He distribution at just a few specific times, and viewing across the poles.

On December 14, 2011, the \lro\ Project altered the spacecraft's polar
orbit from a near-circular 50~km altitude to a more stable, fuel
efficient elliptical orbit with apoapsis near 200~km and periapsis at
30~km.  This change occurred when the orbit was very close to the
terminator (at the terminator the $\beta$-angle, defined as the
difference between the Sun direction and the spacecraft orbit plane, is
90\arcdeg).  As soon as $\beta$ decreased to the point where it was feasible
to observe the dark lunar surface with the full LAMP aperture, we found
that the enhanced He column along the line-of-sight from apoapsis (on the night side the He scale height is about 130~km) made detection of \Heone\ \lam584\ fairly straightforward because of the absence of sky background.  Moreover, it would allow continuous monitoring of the He atmosphere until $\beta$ reached the
point where the spacecraft was totally in shadow, or for roughly a
one-month period.  The technique of observing illuminated atmosphere
above the dark lunar surface is the same as that employed by
\citet{Fastie:1973} with the {\it Apollo 17} Ultraviolet Spectrometer,
except that in that case the near-equatorial orbit permitted only a few
minutes of observation at each terminator crossing.

In this paper we present the details of the spectroscopic analysis of
the additional, more extensive observations made in late December 2011 and January 2012, discuss the calibration issues, and describe the
parameterization used to infer the He surface density from the data.  We will
also discuss our results in terms of the solar wind alpha particle flux
and recent modeling efforts.

\section{Observations}
\label{obslamp}

LAMP is a lightweight, low-power, imaging spectrograph optimized for
far-ultraviolet (FUV) spectroscopy of the nightside lunar surface and
atmosphere \citep{Gladstone:2010b}. It is designed to obtain spatially
resolved spectra in the 570-1960~\AA\ spectral band with a spectral
resolution of 34~\AA\ for extended sources that fill its field-of-view. 
The slit is 6.0\arcdeg\ long, with a width of 0.3\arcdeg.  Each spatial
pixel along the slit is 0.3\arcdeg\ long.  From an altitude of 50~km the
spatial resolution is $0.26 \times 0.26$~km$^2$.  The use of LAMP for
the detection of gaseous emission was demonstrated by the observations
of the plume produced by the LCROSS impact on October 9, 2009
\citep{Gladstone:2010a}.

A diagram showing the observing geometry for the ``frozen'' elliptical
orbit is shown in Fig.~\ref{shadow} for two different dates.  The
initial time in each plot is the equator crossing time on the day side. 
Each orbit, of roughly two hours, generates a separate data file of
time-tagged (pixel list) photon events.  As shown in the cartoon of Fig.~\ref{cartoon}, the orbit crosses the north
pole at roughly t+1800~s, descends to the night side equator and crosses
the south pole near periapsis at $\sim$t+5400~s.  Near the poles there
are bright sunlit spots that cause the detector high voltage to be temporarily
reduced producing data dropouts, so that our useful data range is
typically from t+2200~s to t+5100~s, corresponding to a latitude range
of 80\arcdeg\ N to 80\arcdeg\ S.  For a given day, we co-add the spectra
from all of the orbits to enhance the signal/noise ratio.

In Fig.~\ref{shadow}, the shaded area represents the distribution of
illuminated atmosphere for a nadir-looking line-of-sight.  The shadow
height, the solid line, is calculated for a spherical moon and and is uncertain to the level of a few km.  The dashed line is the \lro\ altitude.  The figure on
the left is for December 29, 2011, with a $\beta$-angle of 83.2\arcdeg. 
The figure on the right is for January 16, 2012, $\beta$=64.7\arcdeg,
and for a good portion of the nightime orbit the spacecraft is in
shadow.  These latter data will prove invaluable in the extraction of
the He emission signal.

\section{Data Analysis}
\label{data}

\subsection{Lunar atmosphere spectrum}
\label{spectrum}

As an example of the short-wavelength region of the LAMP spectrum, Fig.~\ref{spec} shows a composite co-added spectrum from 12 orbits on January 16, 2012, in which the count rate in the illuminated rows of the detector are summed for times between 2100 and 2600~s after the start of the orbit.  This corresponds to the left shaded region in the right panel of Fig.~\ref{shadow}.  The background is due to grating scattered \lya\ photons, from the interplanetary background reflected from the dark surface, and the steep rise to the short-wavelength limit is caused by electronic pile-up at the edge of the sensitive area of the detector.  Another artifact exists at wavelengths 760--800 \AA.  Overplotted in red is a spectrum with the spacecraft totally in shadow, from 3300 to 4700 seconds.  A very small \Heone\ \lam584 signal remains, presumably due to reflected interplanetary \Heone\ \lam584 from the dark lunar surface.  The difference between the two gives the emission from the illuminated column, as seen in the lower panel.  In extracting the He emission as a function of time, the shadow spectrum is used as a template for subtracting the instrumental background.

\subsection{Orbit variation}
\label{orbit}

Fig.~\ref{shadow} suggests that the observed \Heone\ \lam584 emission rate should follow the height of the shaded area as a function of time during the night side of a given orbit.  This behavior is indeed found, as can be seen in Fig.~\ref{counts} for the two dates of Fig.~\ref{shadow}.  To enhance the signal/noise ratio of these data we have co-added all of the orbits for a given date.  [A day-by-day compilation of all of the data for the entire period is available in the Supplementary Online Material, Figs.~\ref{supp1}-\ref{supp3}].  The data are parameterized as described in Section~\ref{model}.

\subsection{Calibration}
\label{cal}

Calibration at 584~\AA\ is difficult.  Although the instrument sensitivity, or effective area, was measured pre-flight in the laboratory, there is no convenient way to monitor the in-flight performnce at this wavelength as is routinely done for wavelengths longward of the interstellar hydrogen absorption edge at 912~\AA\ using a selection of standard stars observed on a regular basis.  Instead, we chose to rely on an observation of strong interplanetary \Heone\ \lam584 emission, made on July 20, 2011, when LAMP was viewing towards the dark limb with the Sun only several degrees below the limb.  The line-of-sight viewed a bright region of the interplanetary He focussing cone.  
We assessed this background and its variation on the sky using a standard helium "hot" model \citep{Ajello:1978} that we have tuned by fitting unpublished {\it Galileo} EUVS interplanetary cruise observations of interplanetary helium obtained at the same time as published observations of interplanetary hydrogen \citep{Pryor:1996}.  Model parameters used for the interstellar helium at large distances from the Sun are a density of 0.015 cm$^{-3}$, velocity of 26.3 km~s$^{-1}$, and a temperature of 6300~K.   The solar \Heone\ \lam584 linewidth is taken as 130 m\AA\ full-width at half-maximum \citep[Doppler width 78 m\AA,][]{Lallement:2004}.    Solar \Heone\ \lam584 brightness values are taken from the Thermosphere Ionosphere Mesosphere Energetics and Dynamics (TIMED) satellite's Solar EUV Experiment (SEE) Level 3 data available online \citep{Woods:1994}.   He atom photoionization rates are taken from the Space Environment Technologies online Solar Irradiance Platform v2.37 \citep{Tobiska:2000}.  We integrated over the slit position on the sky over the five minute duration of the observation and deduced a sensitivity at 584~\AA\ of 0.75~counts s$^{-1}$ rayleigh$^{-1}$.  We have verified consistency by comparing nighttime observations of the sky with the models for those particular dates and view directions.  All of the nadir observations have emission rates of the order of or less than 1~rayleigh.

\subsection{Data Modeling}
\label{model}

We parameterize the He atmosphere as a surface bounded exosphere using the classical model of \citet{Chamberlain:1963}.  \citet{Hartle:1974} have shown this to be a valid approximation for He at low altitudes, such as those considered here.  We take the exobase temperature to be the surface temperature which we set at 100~K \citep{Vasavada:2012}, leaving the helium density at the surface as the sole variable.  We assume that \Heone\ \lam584 is optically thin and calculate a fluorescence efficiency (g-factor) using daily solar \Heone\ \lam584 fluxes from SDO/EVE \citep{Woods:2010} and the solar linewidth quoted above.  The density model is then integrated between the spacecraft orbit and the shadow height to calculate a predicted brightness which is then fit to the data using a linear least-squares algorithm, to infer the He surface density.  The fits are illustrated in Fig.~\ref{counts} and in the Supplementary Online Material Figs.~\ref{supp1}-\ref{supp3}.

For all of our dates, the fit of the model to the data is excellent, and implies that the He surface density is roughly constant with latitude.  Because of the night time He scale height of $\sim$130~km near the dawn terminator, there is also not much sensitivity to the choice of surface temperature, as a variation of $\pm 20$~K gives only a $\pm 9$\% and $\pm 2$\% change in calculated brightness from \lro\ altitudes of 200 and 35~km, respectively, which is within the range of statistical uncertainty in the data fits. 

\subsection{Temporal Variability}
\label{variability}

The variation of the inferred surface He density with date is shown in Fig.~\ref{hedens}.  During this period, the longitude, defined by \citet{Hodges:1973} as the angular distance from the sub-solar meridian, is decreasing.  At the dawn terminator, the longitude is 180\arcdeg $+ |\beta|$, and this is shown on the figure.  This quantity is analogous to local time and is useful in comparing the data with models in the literature.  The mean density inferred for longitudes within 15\arcdeg\ of the terminator is $\sim2.0 \times 10^4$~cm$^{-3}$, in good agreement with the {\it Apollo 17} LACE data taken at a latitude of 20\arcdeg\ for the same range of longitudes.  On a given date, orbit-to-orbit variations are less than the statistical uncertainties in the count rates, so that the time scale for density variability is greater than 2~hours (the orbit period), but of the order of a day, the time scale for thermal escape of He \citep{Hodges:1973}.  Our data show a variation of a factor of three in derived density over a 28~day period.

One striking feature stands out: a decrease in the surface He density by a factor of two in a five-day period beginning on January 7, 2012.  This decrease occurs during a period in which the Moon passes through the Earth's magnetotail.  The shape and magnitude of the decrease match very well the prediction of \citet{Hodges:1978}, and are discussed below.

\section{Discussion}

Following the {\it Apollo 17} mission, discussion and modeling of the
LACE measurements (that were made over 10 consecutive lunations) focussed
on the question of the efficacy and distribution of solar wind alpha
particle neutralization, thermalization, and evaporation as the primary
source of the He exosphere \citep{Hodges:1974}.  Modeling by \citet{Hodges:1973} was able to reproduce the observed longitudinal dependence and magnitude of the LACE measurements, with the assumption that these processes were almost 100\%
efficient. Recent modeling, by \citet{Leblanc:2011}, supports these
conclusions. The temporal variability observed by LAMP would then be
expected to correlate with the solar wind alpha flux.  Similar variations in the night time He density were seen in the LACE data and shown to correlate with the geomagnetic $K_P$ index, again suggesting a dependence on the solar wind flux \citep{Hodges:1974}.  

Using the Monte Carlo lunar exosphere model of \citet{Killen:2012}, we simulate the time-dependent helium density in the lunar exosphere.  
We assume a source of solar wind alpha particles that become neutralized and thermalized through interaction with the lunar regolith on short time scales.  The flux incident on the lunar surface falls off as a function of solar zenith angle.  The aberration in the apparent solar wind direction of 5\arcdeg\ is neglected.  The neutral helium atoms are released into the atmosphere with a velocity taken from a Maxwell-Boltzmann flux distribution \citep{Smith:1978} at the local surface temperature taken from Diviner data \citep{Vasavada:2012}.  
The model tracks the helium atoms as a function of time assuming that the
outbound velocity comes from the local thermal distribution after each
encounter with the surface.  Tracing the evolution of the helium exosphere with time enables a time-dependent source rate to be implemented.

Upstream solar wind monitors can provide measures of the solar wind speed and the alpha particle density at the L1 point.  However, until calibrated data are available for this time period, we assume a constant solar wind alpha flux of 
$1.2 \times 10^7$~cm$^{-2}$~s$^{-1}$, except when the Moon passes through the Earth's magnetotail.  We do not use a different alpha flux for the magnetosheath than for the unshocked solar wind.  Using data from the twin {\it Acceleration,
Reconnection, Turbulence, and Electrodynamics of the Moon's Interaction
with the Sun} ({\it ARTEMIS}) spacecraft in orbit about the Moon
\citep{Halekas:2011}, we identify the times around full moon when the ion flux at the Moon is drastically reduced compared to nominal solar wind.  During these times, we set the source rate to zero.  

In Figure~\ref{hedens}, we show the model density (red diamonds) as a function of time compared to the LAMP observations (black asterisks).  The model density was taken at the lunar surface from the longitude of {\it LRO's} nightside passage for each day.  The trend of increasing density at the beginning of January in the model is the result of the progression of {\it LRO's} orbit further from the terminator.  Deviations between the model density and the LAMP observations are probably due to short-term variations in the solar wind alpha flux.  The decrease in He density beginning on January 7, however, does not appear to correlate with a comparable decrease in solar wind flux.  As noted above, this time corresponds to the Moon's path entering into Earth's magnetotail which shields the surface of the Moon from the outflowing solar wind.  During this time the He density appears to decrease exponentially with a time constant of $\sim$5~days which we associate with the thermal loss rate of He atoms at the corresponding surface temperature.  Once the Moon exits from the magnetotail, the density returns to its former value with a similar time constant.  Our model closely tracks the temporal behavior of the inferred He surface density and is also in qualitative agreement with the initial predictions of  \citet{Hodges:1978}.  The LAMP data are the first such measurements to confirm these predictions.

We note that a large solar coronal mass ejection (CME) occurred on January 23-24 and produced an extremely large increase in solar wind flux but also generated a large flux of MeV electrons that
produced a highly elevated background count rate on the LAMP detector
and masked any \Heone\ \lam584 signal.  After the event, the inferred He
density does not appear higher than before the event, but by this time
the $\beta$-angle had decreased to the point where only a small fraction
of the orbit had illuminated atmosphere and our measurements are limited
to high latitudes.

One caveat about our measurement technique is that it does not give us
access to the complete range of longitudes and latitudes on the night
side of the Moon.  We are limited to range of about 30\arcdeg\ in
longitude on the night side of both the dusk and dawn terminators.  However, we expect to be able to measure the dawn/dusk asymmetry in He density seen in
the LACE data, as well as to track long term correlations with the solar wind alpha particle flux.

\section{Conclusion}

We have described continuous observations of the night side lunar helium
exosphere made over a period of a month in January 2012 using the Lyman
Alpha Mapping Project ultraviolet spectrograph on NASA's \lro\ mission.
These observations show that the inferred surface He density exhibits
day-to-day variations that likely vary with the solar wind alpha flux
and decrease significantly during the passing of the Moon through the Earth's magnetotail.  Once calibrated solar wind alpha particle fluxes measured by the Solar Wind Electron Proton Alpha Monitor (SWEPAM) instrument on the {\it Advanced Composition Explorer (ACE)} satellite are publicly available for January 2012, we intend to carry out a more detailed investigation of the correlation between the solar wind fluxes and inferred surface He density.  We also expect additional LAMP data from dawn and dusk terminator crossings during the summer of 2012, which we will report in a future publication.

\begin{center}{\bf Acknowledgments}\end{center}

We thank the {\it Lunar Reconnaissance Orbiter} project and project team at
NASA's Goddard Space Flight Center for their continuous support.  We thank Jasper Halekas for guidance with {\it ARTEMIS} data.  We acknowledge
NASA contract NAS5-02099 and V. Angelopoulos for use of data from the
{\it THEMIS/ARTEMIS} mission, and thank specifically C. W. Carlson and J. P. McFadden for use of ESA data.  This work was financially supported under contract NNG05EC87C from NASA to the Southwest Research Institute.  The work at Johns Hopkins University was supported by a sub-contract from Southwest
Research Institute.

\clearpage

\renewcommand\baselinestretch{1.2}%

\clearpage

\clearpage

\renewcommand\baselinestretch{1.6}%

\clearpage 
\begin{center}{\bf FIGURE CAPTIONS}\end{center}

\noindent
Fig.~\ref{shadow}.  \lro\ altitude (dashed line) and calculated shadow height (solid line) for the first orbit on December 30, 2011 (left) and January 16, 2012 (right).  The origin of the time axis is the time of equator crossing on the day side of the orbit.  The shaded area shows the region of solar illuminated atmosphere below \lro.

\vspace*{0.2in}

\noindent
Fig.~\ref{cartoon}.  Cartoon showing the elliptical \lro\ polar orbit (in red) relative to the dawn terminator and the $\sim$130~km He scale height.  Apoapsis is near the north pole (at top), the Sun is to the right.  LAMP, viewing towards the nadir, observes illuminated atmosphere against the dark surface of the Moon.

\vspace*{0.2in}

\noindent
Fig.~\ref{spec}.  Average of 12 short wavelength LAMP spectra of the illuminated (black) and dark (red) atmosphere on January 16, 2012.  The feature centered at 775~\AA\ is an instrumental artifact.  The lower panel is the difference, the average atmospheric component of He emission.

\vspace*{0.2in}

\noindent
Fig.~\ref{counts}.  Observed count rates for the atmospheric \Heone\ \lam584 emission for the two dates of Fig.~\ref{shadow}.  For both dates the data from 12 contiguous orbits have been co-added.  The error bars are 1-$\sigma$ statistical uncertainties.  The overplotted models are discussed in the text.

\vspace*{0.2in}

\noindent
Fig.~\ref{hedens}.  Surface density of helium is shown as a function of time from LAMP observations (black asterisks) and model predictions (red diamonds)
averaged over each day.  Error bars denote estimated 1-$\sigma$ uncertainties in the fit of the data to the exospheric column density model.  The progression in longitude of LRO's nightside groundtrack is provided near the bottom.  Times of the Moon's passage through the Earth's magnetotail are indicated by dashed lines.

\newpage

\renewcommand\baselinestretch{1.2}%
\setcounter{figure}{0}

\begin{figure*}[ht]
\begin{center}
\includegraphics*[width=0.49\textwidth,angle=0.]{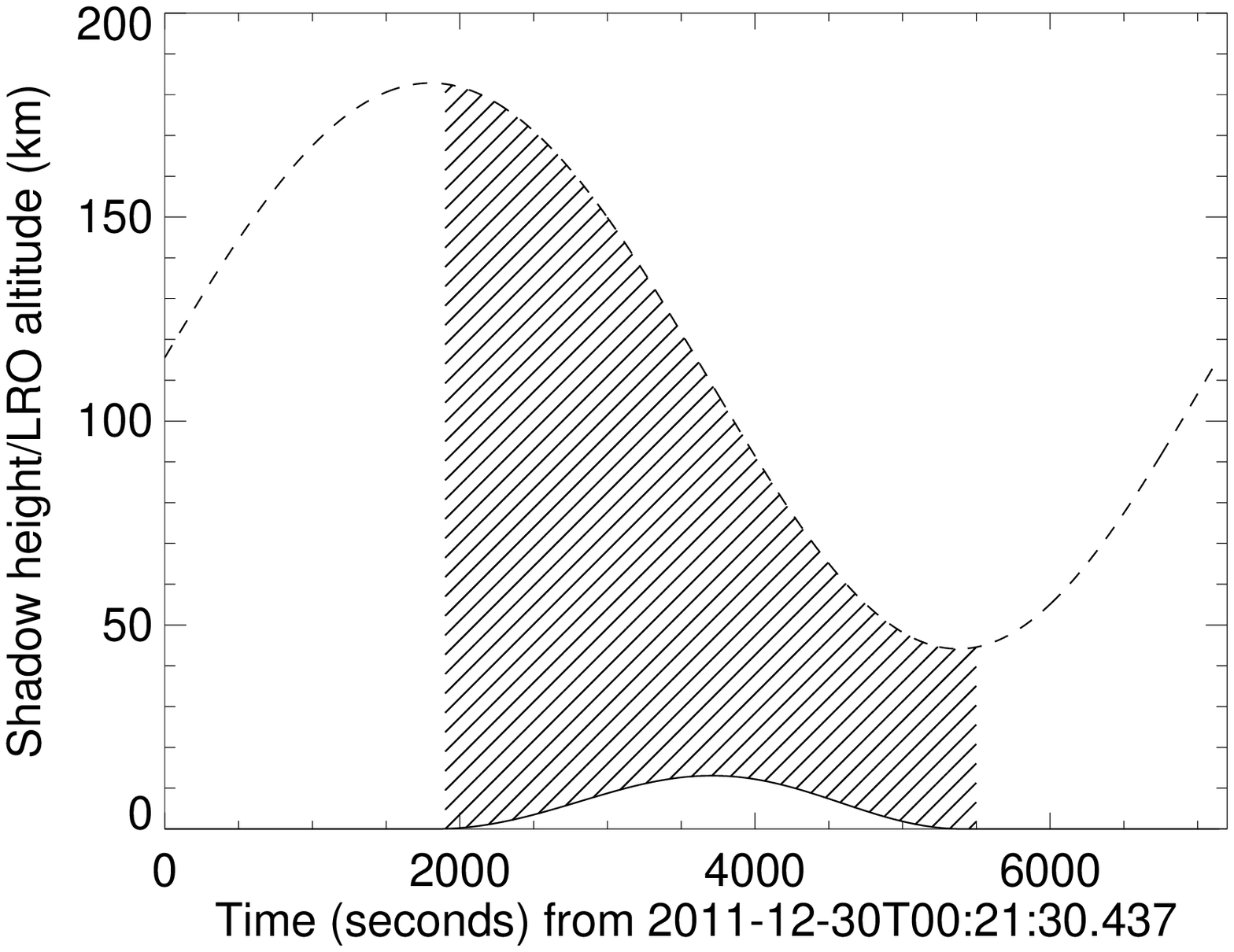}
\hfill
\includegraphics*[width=0.49\textwidth,angle=0.]{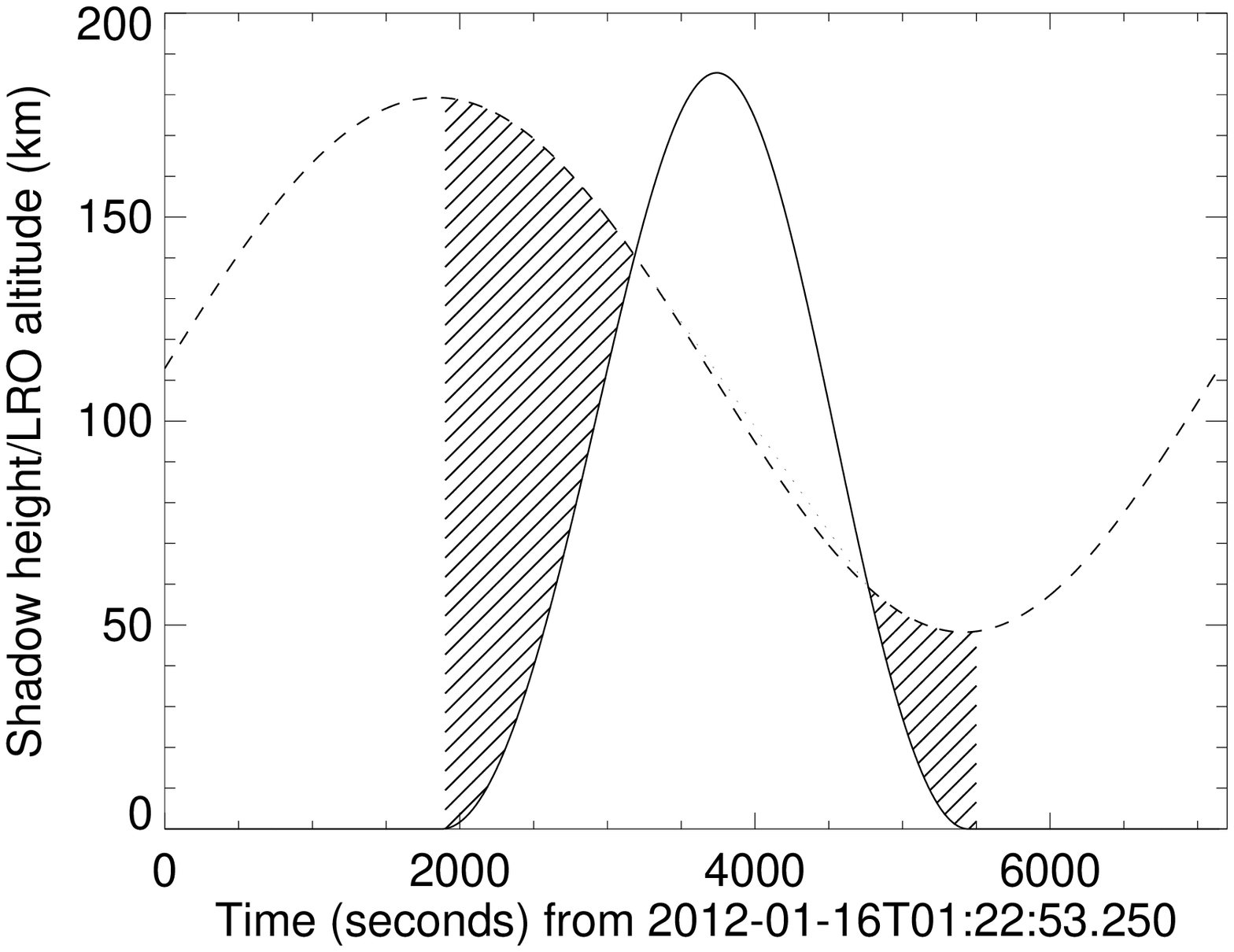}
\caption[]{\lro\ altitude (dashed line) and calculated shadow height (solid line) for the first orbit on December 30, 2011 (left) and January 16, 2012 (right).  The origin of the time axis is the time of equator crossing on the day side of the orbit.  The shaded area shows the region of solar illuminated atmosphere below \lro. \label{shadow} }
\end{center}
\end{figure*} 

\begin{figure*}[ht]
\begin{center}
\includegraphics*[width=0.75\textwidth,angle=0.]{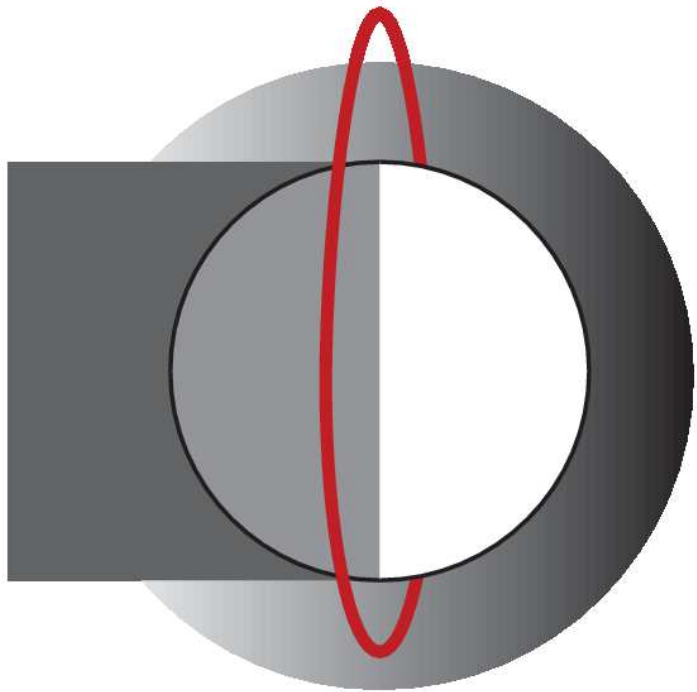}
\vspace*{-0.2in}
\caption[]{Cartoon showing the elliptical \lro\ polar orbit (in red) relative to the dawn terminator and the $\sim$130~km He scale height.  Apoapsis is near the north pole (at top), the Sun is to the right.  LAMP, viewing towards the nadir, observes illuminated atmosphere against the dark surface of the Moon.  \label{cartoon} }
\end{center}
\end{figure*} 

\begin{figure*}[ht]
\begin{center}
\includegraphics*[width=0.9\textwidth,angle=0.]{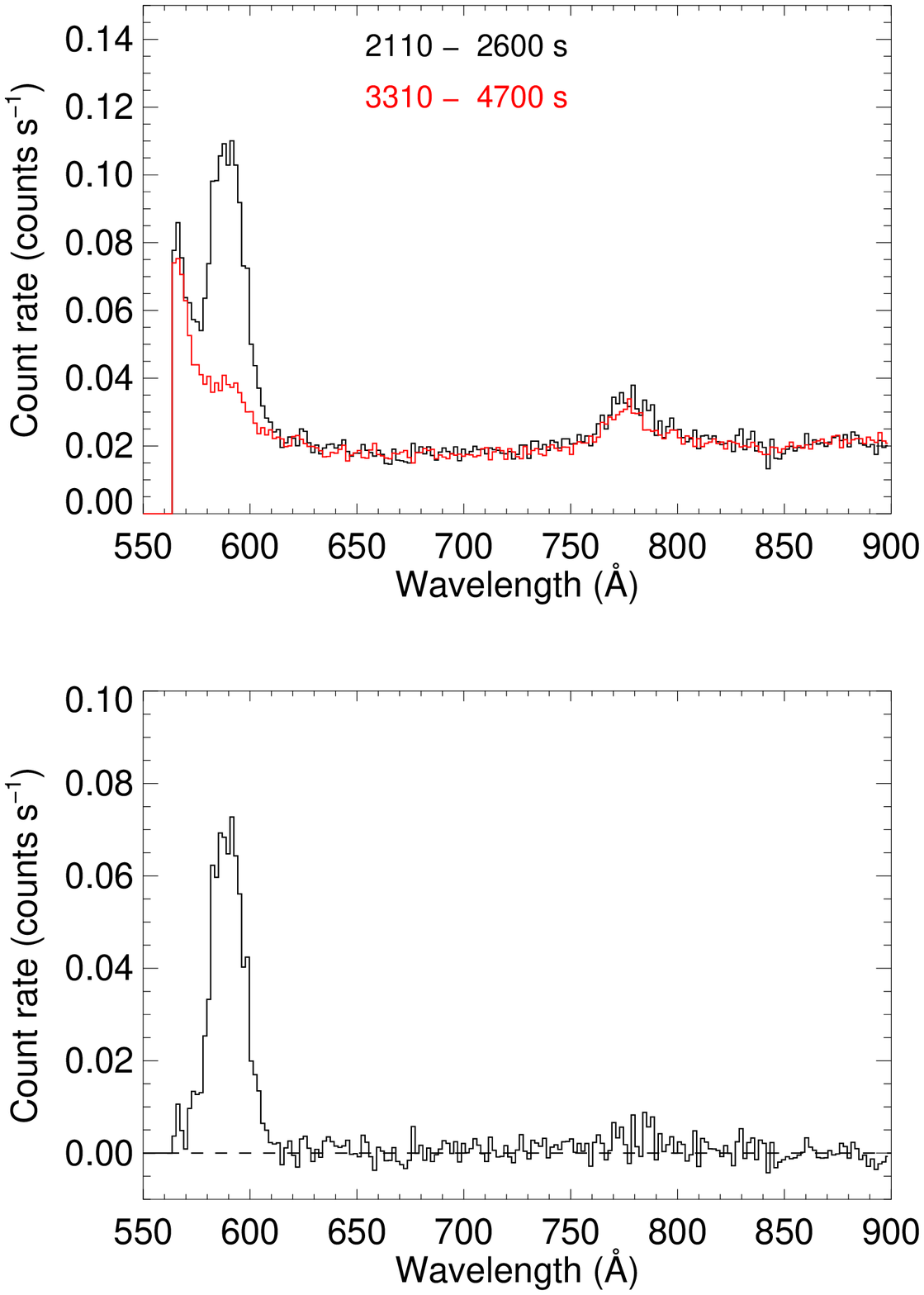}
\vspace*{-0.2in}
\caption[]{Average of 12 short wavelength LAMP spectra of the illuminated (black) and dark (red) atmosphere on January 16, 2012.  The feature centered at 775~\AA\ is an instrumental artifact.  The lower panel is the difference, the average atmospheric component of He emission.  \label{spec} }
\end{center}
\end{figure*} 

\begin{figure*}[ht]
\begin{center}
\includegraphics*[width=0.49\textwidth,angle=0.]{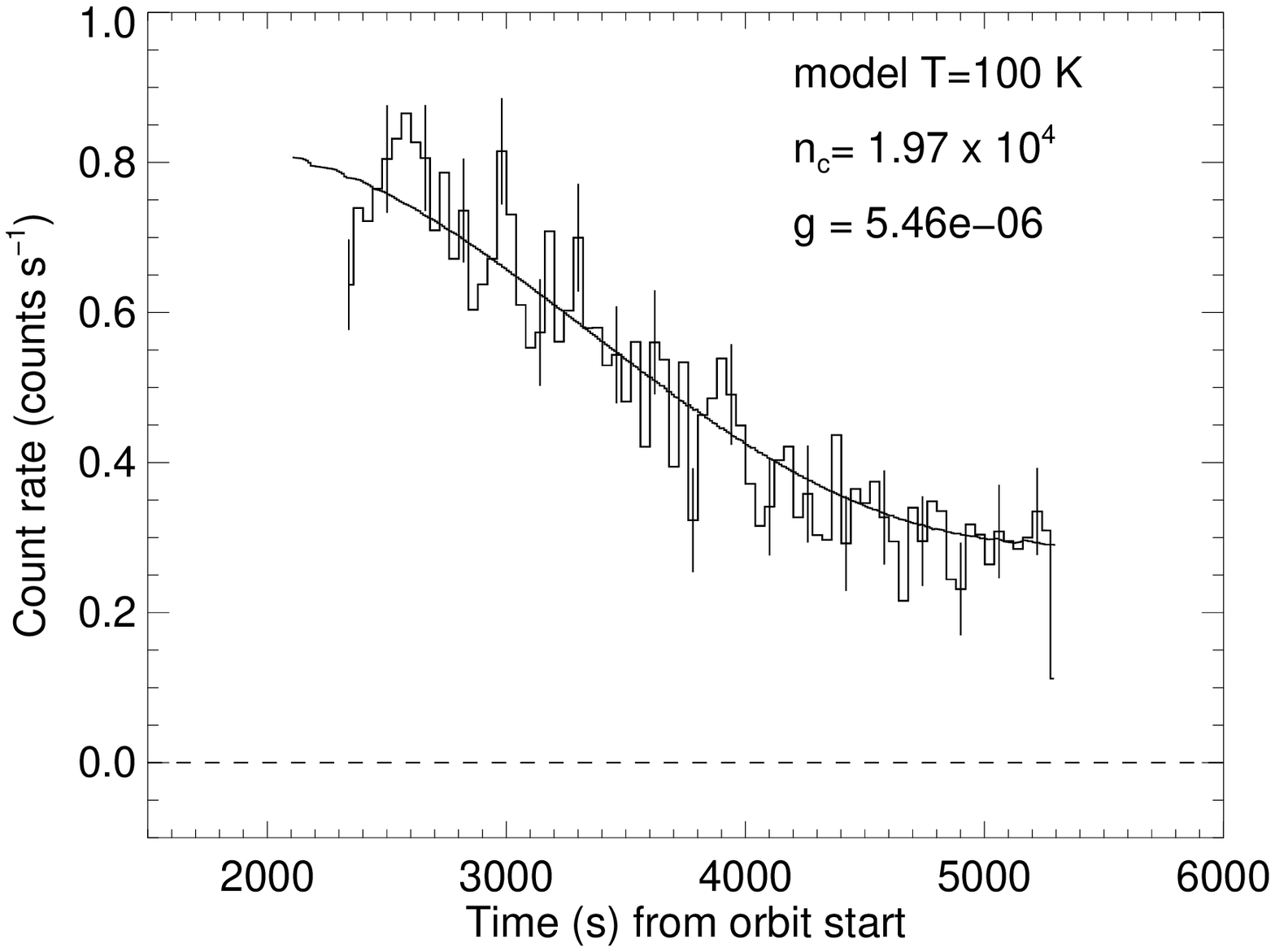}
\hfill
\includegraphics*[width=0.49\textwidth,angle=0.]{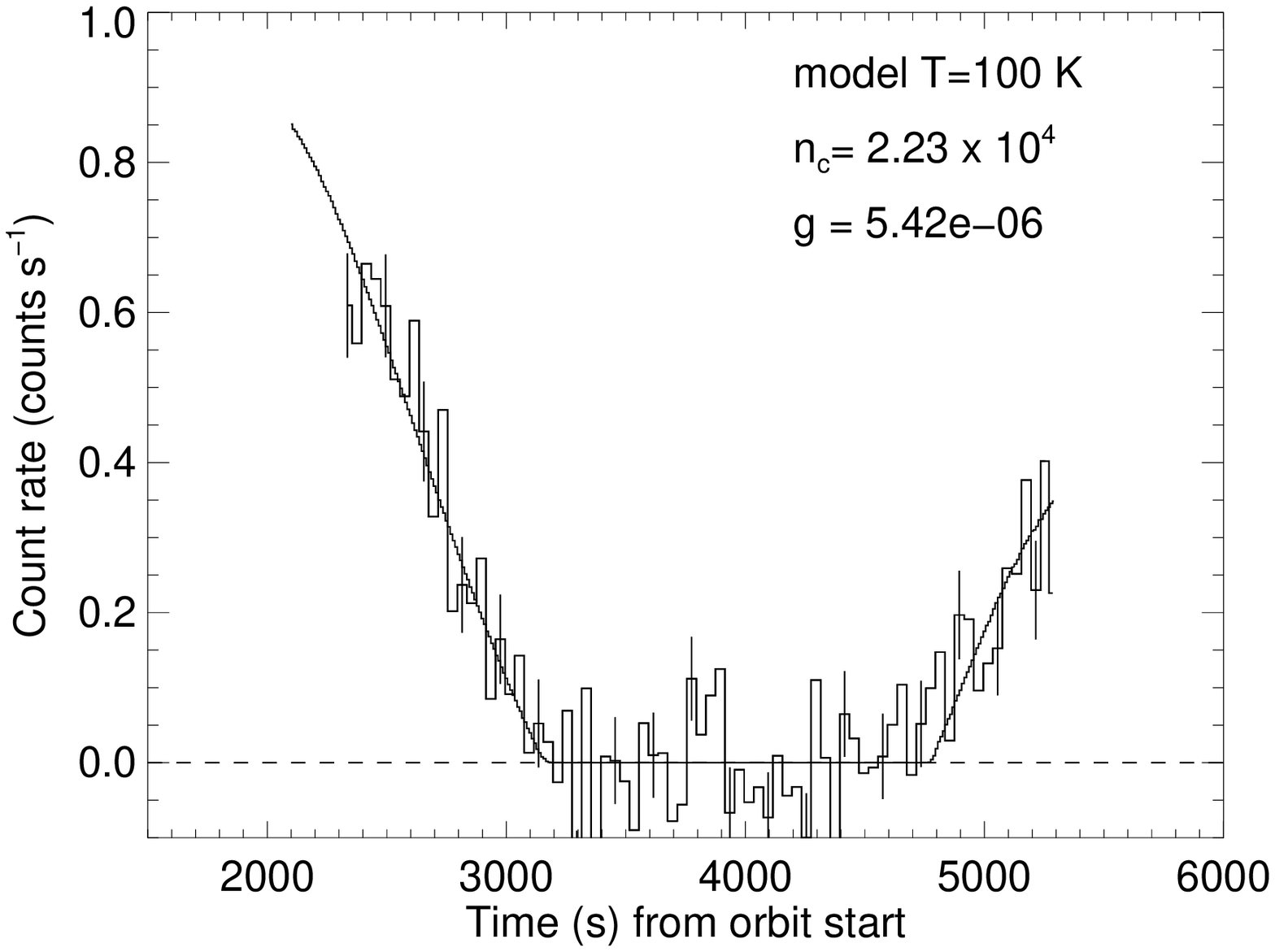}
\caption[]{Observed count rates for the atmospheric \Heone\ \lam584 emission for the two dates of Fig.~\ref{shadow}.  For both dates the data from 12 contiguous orbits have been co-added.  The error bars are 1-$\sigma$ statistical uncertainties.  The overplotted models are discussed in the text. \label{counts} }
\end{center}
\end{figure*} 

\begin{figure*}[ht]
\begin{center}
\includegraphics*[width=1.0\textwidth,angle=0.]{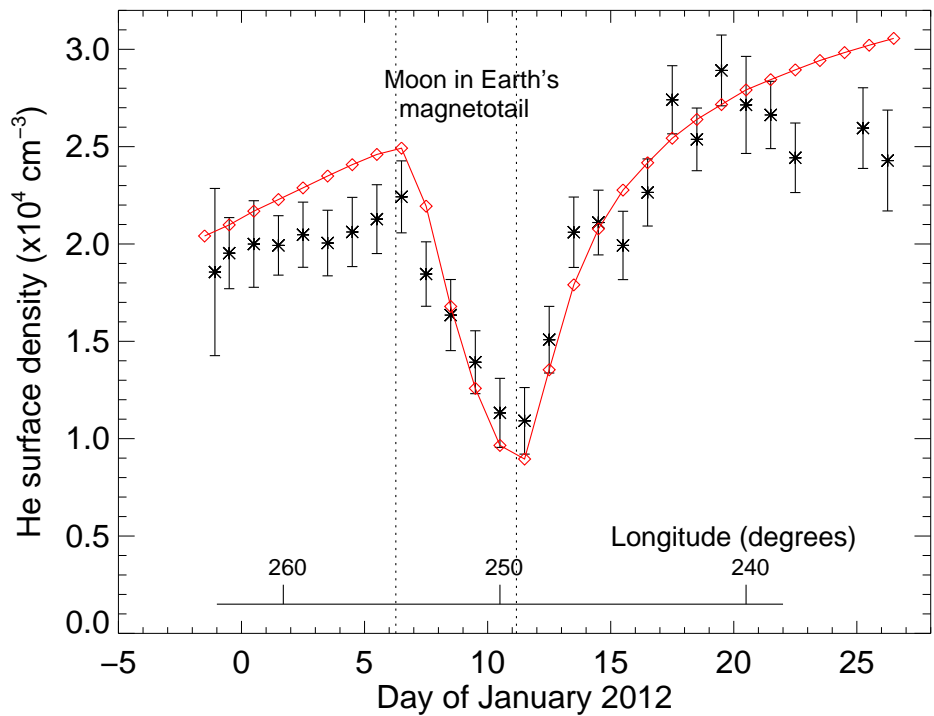}
\vspace*{-0.2in}
\caption[]{Surface density of helium is shown as a function of time from LAMP
observations (black asterisks) and model predictions (red diamonds)
averaged over each day.  Error bars denote estimated 1-$\sigma$ uncertainties in the fit of the data to the exospheric column density model.
The progression in longitude of LRO's nightside groundtrack is provided
near the bottom.  Times of the Moon's passage through the Earth's
magnetotail are indicated by dashed lines.  \label{hedens} }
\end{center}
\end{figure*} 

\setcounter{figure}{5}
\clearpage 
\begin{center}{\bf SUPPLEMENTARY ONLINE MATERIAL}\end{center}

\noindent
Figures~6-8.  The following plots show the same data as in Fig.~\ref{counts}, but day-by-day from December 30, 2011 through January 22, 2012.  For each date the number of contiguous orbits that have been co-added is indicated in the title, together with the day of year (DOY) and year.  The error bars are 1-$\sigma$ statistical uncertainties.  The parameters used to fit the overplotted models are also indicated on each plot.

\vspace*{0.3in}

\begin{figure*}[ht]
\begin{center}
\includegraphics*[width=0.9\textwidth,angle=0.]{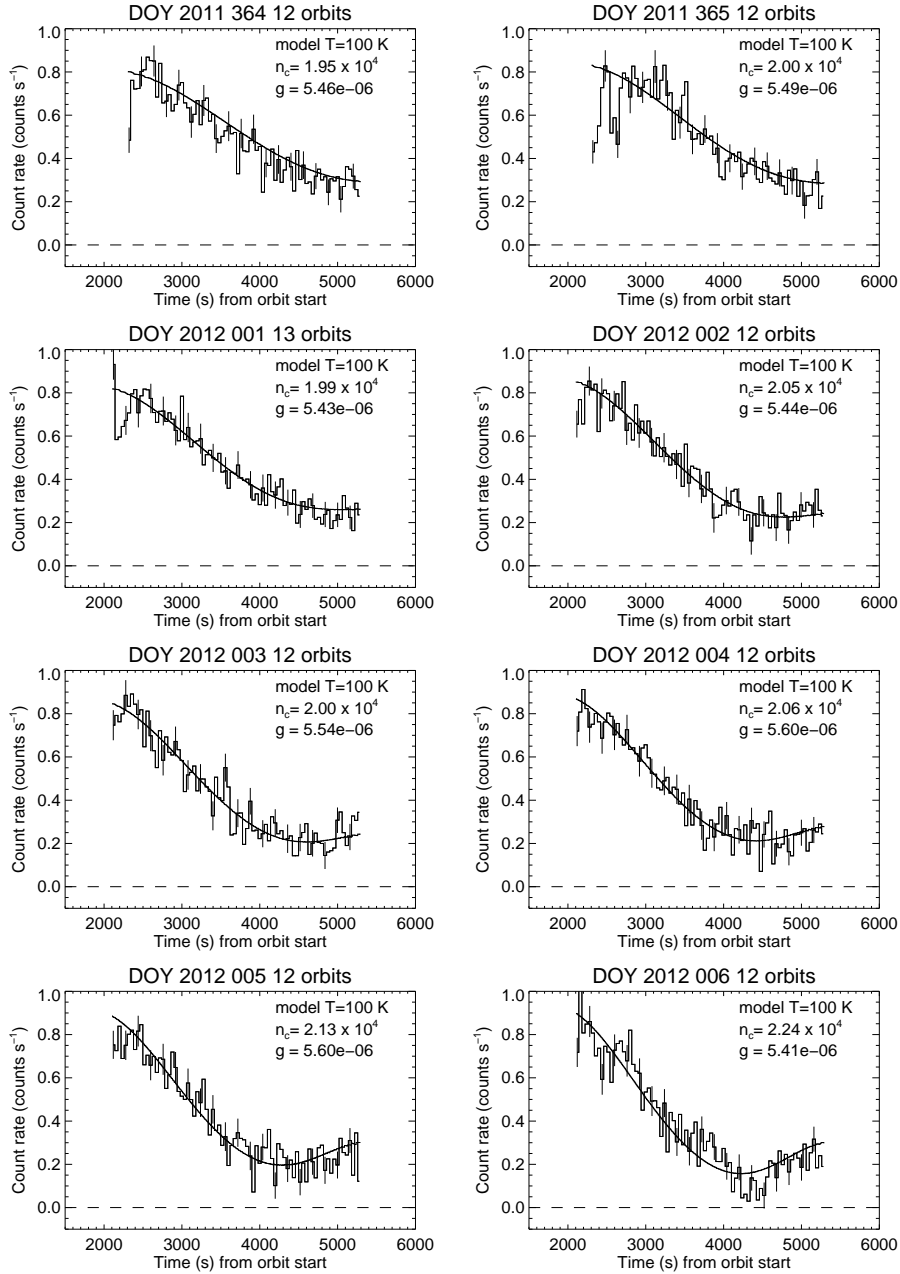}
\caption[]{December 30, 2011 to January 6, 2012. \label{supp1} }
\end{center}
\end{figure*}

\noindent
Fig.~\ref{supp1}.  December 30, 2011 to January 6, 2012.

\vspace*{0.2in}

\begin{figure*}[ht]
\begin{center}
\includegraphics*[width=0.9\textwidth,angle=0.]{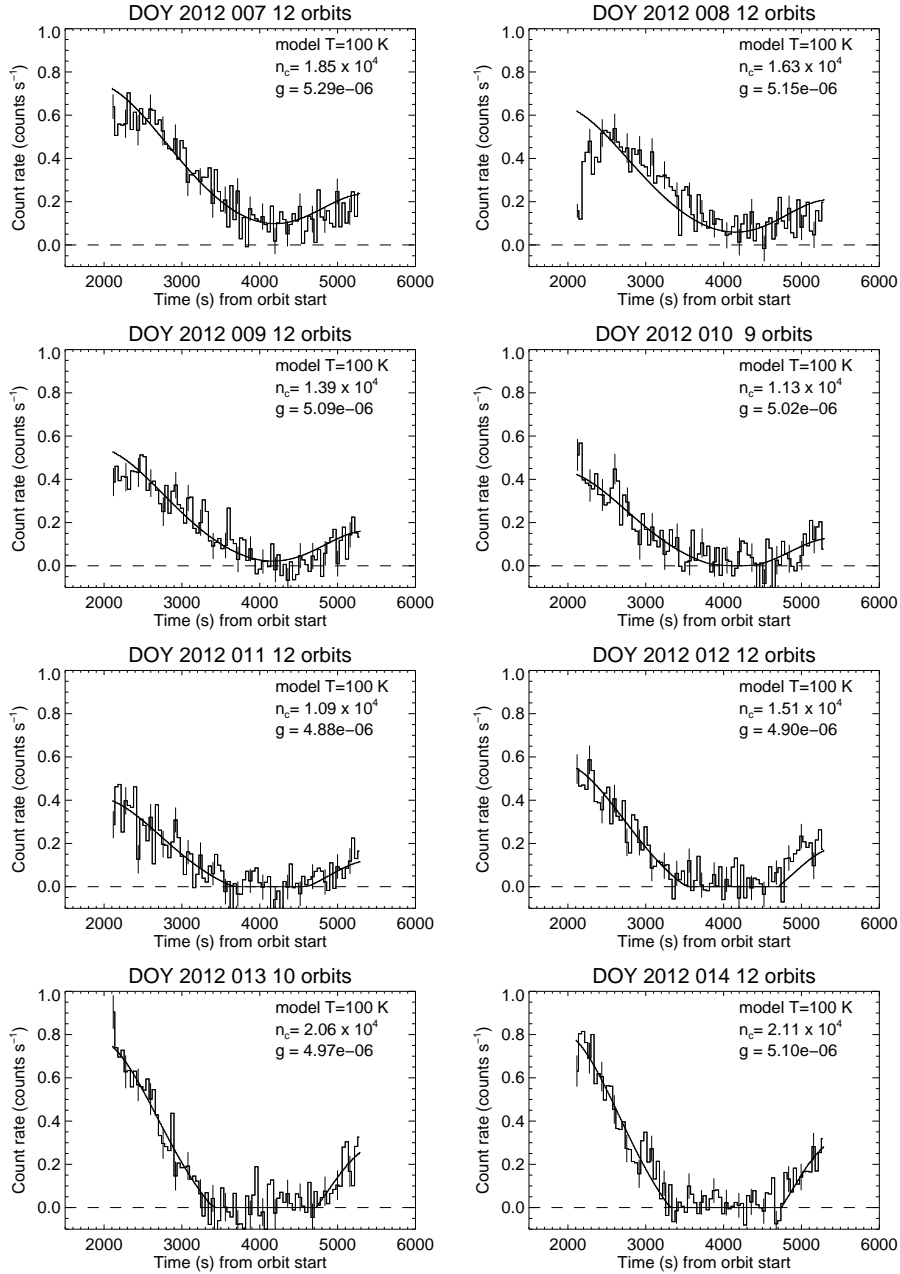}
\caption[]{January 7, 2012 to January 14, 2012. \label{supp2} }
\end{center}
\end{figure*}

\noindent
Fig.~\ref{supp2}.  January 7, 2012 to January 14, 2012.

\vspace*{0.2in}

\begin{figure*}[ht]
\begin{center}
\includegraphics*[width=0.9\textwidth,angle=0.]{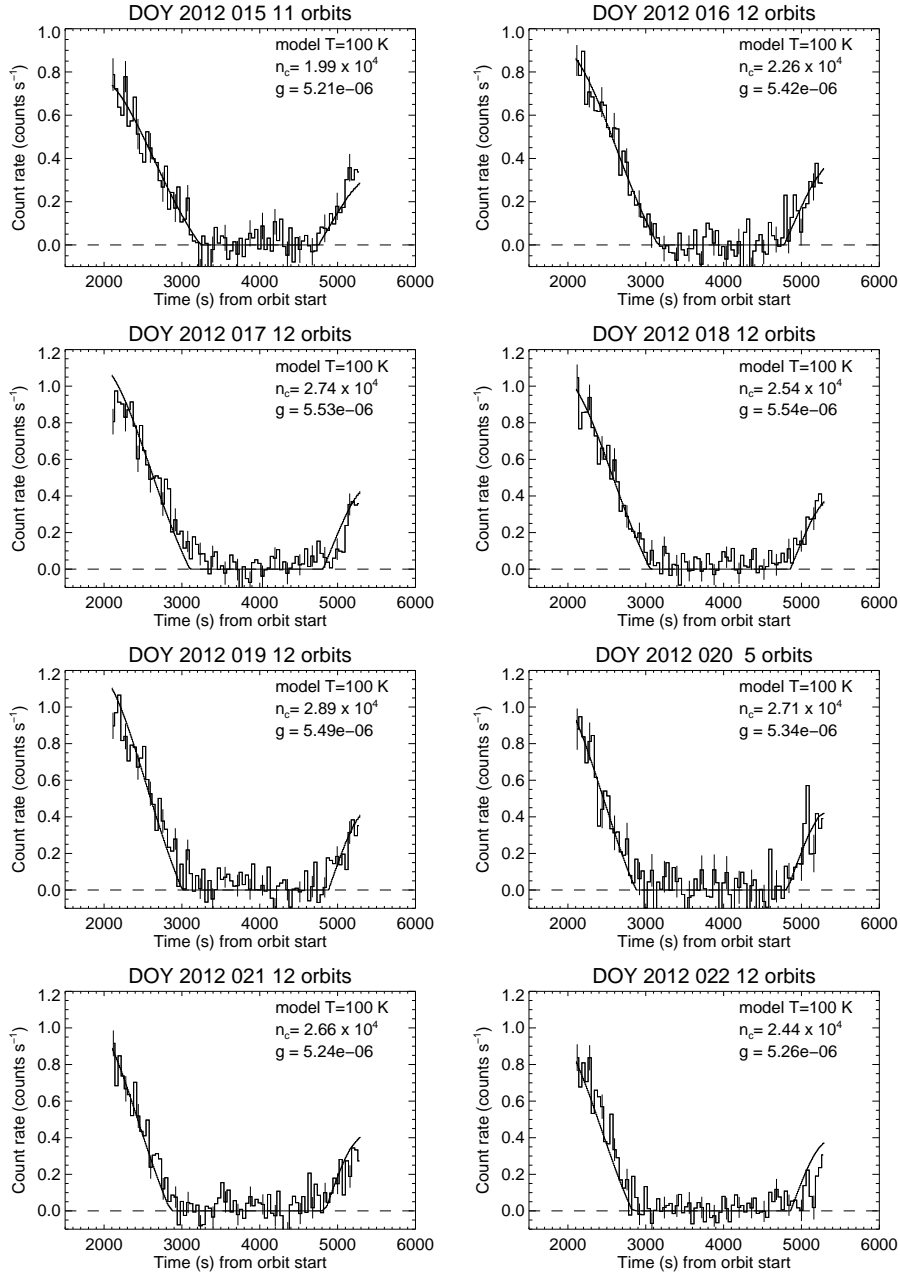}
\caption[]{January 15, 2012 to January 22, 2012. \label{supp3} }
\end{center}
\end{figure*}

\noindent
Fig.~\ref{supp3}.  January 15, 2012 to January 22, 2012.

\end{document}